# Solitons in a forced nonlinear Schrödinger equation with the pseudo-Raman effect

**Evgeny M. Gromov[1] and Boris A. Malomed[2]**

[1]National Research University Higher School of Economics (HSE), Nizhny Novgorod 603155, Russia

[2]Department of Physical Electronics, School of Electrical Engineering, Faculty of Engineering, Tel Aviv University, Tel Aviv 69978, Israel

**Abstract**

Dynamics of solitons is considered in the framework of an extended nonlinear Schrödinger equation (NLSE), which is derived from a Zakharov-type model for wind-driven high-frequency (HF) surface waves in the ocean, coupled to damped low-frequency (LF) internal waves. The drive gives rise to a convective (but not absolute) instability in the system. The resulting NLSE includes a *pseudo-stimulated-Raman-scattering* (pseudo-SRS) term, which is a spatial-domain counterpart of the SRS term, a well-known ingredient of the temporal-domain NLSE in optics. Analysis of the field-momentum balance and direct simulations demonstrate that wavenumber downshift by the pseudo-SRS may be compensated by the upshift induced by the wind traction, thus maintaining robust bright solitons in both stationary and oscillatory forms; in particular, they are not destroyed by the underlying convective instability. Analytical soliton solutions are found in an approximate form and verified by numerical simulations. Solutions for soliton pairs are obtained in the numerical form.

**Keywords:** Zakharov system, extended nonlinear Schrödinger equation, soliton, induced scattering, damped low-frequency waves, wind pump, method of moments, numerical simulations

**PACS:** 05.45.Yv; 52.35.Sb; 42.65.Tg; 42.65.Dr

## 1. Introduction

The ability of solitons to keep robust shapes while traveling long distances may be harnessed for the stable transfer of energy, matter, and signals, making solitons an important object of research in diverse areas. Soliton solutions emerge in a broad class of models dealing with the propagation of waves in dispersive nonlinear media, such as surface waves on deep water, light pulses and beams in photonic media, electromagnetic waves in plasma, etc. [1-8]. More recent realizations of solitons were found in the form of self-trapped matter-wave pulses in various forms of Bose-Einstein condensates [9], and as localized excitations in plasmonics [10].

In the above-mentioned contexts, solitons appear as self-trapped packets carried by high-frequency (HF) waves. The HF dynamics is governed by the second-order nonlinear dispersive wave theory, whose fundamental equation is the nonlinear Schrödinger equation (NLSE) with the second-order dispersion (SOD) and cubic self-phase modulation [11,12]. As is commonly known, stable soliton solutions arise, in this case, as a result of the balance between the dispersive self-stretching and nonlinear self-compression of wave packets. Additional balance conditions may support robust solitons in media featuring dissipation and spatial inhomogeneity. In particular, permanent-shape solutions for solitons were found in the framework of the NLSE including linear losses of HF waves and spatially-decreasing SOD [13,4].

The dynamics of short intense HF wave packets is described by the third-order nonlinear dispersive wave theory [1], which takes into account the third-order dispersion (TOD), nonlinear dispersion (self-steeping) [16], and stimulated Raman scattering (SRS), in the case of optical fibers [17-19]. The basic equation of the theory is an extended NLSE [19-23]. Soliton solutions in the framework of the extended NLSE including the TOD and nonlinear dispersion were found in Refs. [24-30]. Stationary shock-wave states, supported by the balance between the SRS and nonlinear dispersion terms, were found as solutions of the extended NLSE in Refs. [31,32]. For solitons, the SRS gives rise to the downshift of the soliton spectrum [17-20], and eventually to destabilization of the solitons. However, a possibility to use the balance between the SRS and gain slope for the stabilization of solitons in long telecommunication links was demonstrated in Ref. [33]. Further, the compensation of the SRS by emission of linear radiation waves from the soliton's core was considered in Ref. [34]. In addition, the compensation of the SRS in inhomogeneous media was considered in several situations, *viz.*, periodically modulated SOD [35], shifting zero-dispersion point [36], and dispersion-decreasing fibers [37].

Proceeding to other physical media, intense short pulses of HF surface waves on deep water, as well as HF Langmuir wave in plasmas, suffer effective damping due to scattering on low-frequency (LF) waves, which, in turn, are subject to the action of viscosity. These LF modes are internal waves in the stratified fluid, or ion-sound waves in the plasma. A model for the HF damping, induced by the coupling to the viscosity-affected LF waves, was proposed in Ref. [38]. It gives rise to an extended NLSE with the spatial-domain counterpart of the SRS term, that was call a *pseudo-SRS* one. The equation was derived from the system of the Zakharov's type [39] for the coupled HF and LF waves. The pseudo-SRS leads to the self-wavenumber downshift, similar to the above-mentioned Raman-induced frequency shift in fiber optics. The model elaborated in Ref. [38] also included smooth spatial variation of the SOD, accounted for by a spatially decreasing SOD coefficient, which leads to an increase of the soliton's wavenumber, making it possible to compensate the effect of the pseudo-SRS on the soliton by the spatially inhomogeneous SOD.

In addition to many well-known models of the nonlinear-wave propagation in optics [4,40,41], in which losses are compensated by gain, a Zakharov-type system for wind-driven surface waves in the ocean, coupled to damped internal-wave modes, was recently introduced in Refs. [14,15]. As shown below, the system may be reduced to an NLSE that includes the pseudo-SRS effect in combination with a linear forcing term, which makes the model subject to *convective* (but not absolute) instability. The objective of the present analysis is to demonstrate that the balance between the pseudo-SRS-induced downshift and forcing-induced upshift of the wavenumber gives rise to a family of *stable* solitons, in spite of the presence of the underlying convective instability. Soliton solutions are found in an explicit approximate form, and verified by dint of direct simulations. Previously, immunity to transverse convective instabilities was shown for quasi-one-dimensional dark solitons in Bose-Einstein condensates [42,43]. To the best of our knowledge, the present analysis demonstrates a similar effect for the bright solitons for the first time.

The model and the analytical approximation, based on moment equations, are introduced in Section II, where estimate for characteristic physical parameters are given too. Analytical and numerical results are reported in Sections III and IV, respectively. The paper is concluded by Section V.

## II. The basic equations, physical estimates, and integral relations

We consider the unidirectional copropagation of a slowly varying envelope, $U(x,t)$, of the complex HF wave field, $U(x,t)\exp(ik_0 x - i\omega_0 t)$, and its real LF counterpart, $n(x,t)$ (effectively, it is a local perturbation of the refractive index). If the HF and LF fields represent the surface and internal waves (SW and IW, respectively) in the ocean, the corresponding system of the Zakharov-type equations is composed of the Schrödinger equation for the SW and Boussinesq equation for the IW, coupled by the usual (quadratic) terms [44,45]. Although the underlying geometry of the fluid motion is two-dimensional, the derivation of the coupled system simplifies the model to the one-dimensional form, as the crucially important geometric elements which guide the propagating waves, *viz.*, the free surface and interface between the layers with different densities of water, are one-dimensional. Under the commonly adopted assumption of the unidirectional wave propagation, the Boussinesq equation may be reduced to one of the Korteweg - de Vries type. Taking into regard LF viscosity $\delta$ and the linear gain with real coefficient $\beta$ applied to the SW, which, as said above, represents the wind forcing in the ocean [15], the system of equations take the form of:

$$2i\left(\frac{\partial U}{\partial t} + V\frac{\partial U}{\partial x}\right) - \frac{\partial^2 U}{\partial x^2} - \beta\frac{\partial U}{\partial x} - nU = 0, \tag{1a}$$

$$\frac{\partial n}{\partial t} + V_L\frac{\partial n}{\partial x} - \delta\frac{\partial^2 n}{\partial x^2} = -\frac{\partial(|U|^2)}{\partial x}, \tag{1b}$$

where $V$ and $V_L$ are the HF and LF group velocities.

The interplay of the wind, SW and IW will be strong enough if the group velocities of the SW and IW at some (widely different, see below) wavelengths, $\Lambda_{\mathrm{SW}}$ and $\Lambda_{\mathrm{IW}}$, are in resonance, and, additionally, the wind's friction velocity, $W$, is in resonance with the SW group velocity [14,45]. Taking a characteristic value, $W \sim 10$ cm/s [46], the classical dispersion relation for the SW on deep water, $\omega_{\mathrm{SW}} = \sqrt{gk}$, and the characteristic value for the Brunt-Väisälä (buoyancy) frequency, $\omega_{\mathrm{BV}} \sim 0.01$ Hz, which gives rise to the IW at the interface between the top mixed layer and the underlying undisturbed one in the ocean (at the depth of a few hundred meters) [47], one can conclude that the corresponding characteristic HF is $\omega_{\mathrm{SW}} \sim 50$ Hz, which exceeds the above-mentioned LF, $\omega_{\mathrm{BV}}$ by three or four orders of magnitude, thus completely justifying the HF-LF distinction. The difference in the respective wavelength is dramatic too, the estimate yielding $\Lambda_{\mathrm{SW}} \sim 2$ cm and $\Lambda_{\mathrm{IW}} \sim 10$ m.

The lowest approximation of the nonlinear dispersion-wave theory corresponds to replacing Eq. (1b) by the adiabatic approximation, $n = |U|^2(V - V_L)^{-1}$, hence envelope $U$ of the HF wave packet obeys the forced NSLE with the linear-gain term [15]:

$$2i\frac{\partial U}{\partial t} = \frac{\partial^2 U}{\partial \xi^2} + \beta\frac{\partial U}{\partial \xi} + 2\alpha U|U|^2, \tag{2}$$

where $\xi \equiv x - Vt$, $\alpha \equiv (1/2)(V - V_L)^{-1}$. At the next order, which takes into regard a correction to the adiabatic response of the LF field, $n = |U|^2 (V - V_L)^{-1} - \delta(V - V_L)^{-2} \partial(|U|^2)/\partial \xi$, Eq. (2) is supplemented by the additional term, which represents the pseudo-SRS effect [38]:

$$2i \frac{\partial U}{\partial t} = \frac{\partial^2 U}{\partial \xi^2} + \beta \frac{\partial U}{\partial \xi} + 2\alpha U |U|^2 - \mu U \frac{\partial(|U|^2)}{\partial \xi}, \qquad (3)$$

where $\mu \equiv \delta(V_L - V)^{-2}$. Below, we fix $\alpha = 1$ by means of obvious scaling.

The gain term in Eq. (3) may be formally absorbed by a transition into a reference frame moving with imaginary velocity, i.e., replacement of real coordinate $\xi$ by $\Xi \equiv \xi - i(\beta/2)t$, which makes it possible to obtain exact soliton solutions to Eq. (3) that explicitly feature growth effects induced by the gain [15]. Here, we prefer to consider Eq. (3) in terms of the real coordinate and time. Then, it is natural to analyze the dispersion relation for small-amplitude excitations, governed by the linearized versions of Eq. (3), by substituting $U \sim \exp(i\kappa\xi - i\omega t)$, which produces a complex frequency as a function of real wavenumber $\kappa$:

$$\omega = -\kappa^2 / 2 + (i/2)\beta\kappa.$$

The same branch of the HF dispersion relation is valid for system (1), as the nonlinear HF-LF coupling does not affect the dispersion relation. The real part of the frequency gives rise to the group velocity, $V_{gr} \equiv d\omega/d\kappa = -\kappa$, hence the excitation traveling at this velocity grows with the distance, $-\xi$, as

$$U \sim \exp(\mathrm{Im}\,\omega \cdot t) \equiv \exp(\mathrm{Im}\,\omega \cdot \xi / V_{gr}) = \exp(-\beta\xi/2) \qquad (4)$$

(note that it does not depend on the wavenumber, $\kappa$), which represents a typical manifestation of the *convective instability* [48]. This type of the instability implies that (in contrast with the *absolute instability*, which drives the growth of quiescent perturbations), the perturbations grow as they travel away, hence they usually do not destroy the underlying patterns. Namely, if a soliton of size $L$, maintained by the balance between the linear gain and pseudo-SRS term, does not move on the average (see below), it follows from Eq. (4) that the soliton is not hurt by the convective instability, provided that it is narrow enough, $L << \beta^{-1}$.

Equation (3) with zero boundary conditions at infinity, $U|_{\xi \to \pm\infty} \to 0$, gives rise to the following integral relations for field moments:

$$\frac{dN}{dt} \equiv \frac{d}{dt} \int_{-\infty}^{+\infty} |U|^2 \, d\xi = \beta \int_{-\infty}^{+\infty} k|U|^2 \, d\xi \equiv -\beta P, \qquad (5)$$

$$\frac{dP}{dt} = -\beta \int_{-\infty}^{+\infty} \left|\frac{\partial U}{\partial \xi}\right|^2 d\xi + \frac{\mu}{2} \int_{-\infty}^{+\infty} \left[\frac{\partial(|U|^2)}{\partial \xi}\right]^2 d\xi, \qquad (6)$$

$$\frac{d}{dt} \int_{-\infty}^{+\infty} \xi |U|^2 \, d\xi = P + \beta \int_{-\infty}^{+\infty} k\xi |U|^2 \, d\xi, \qquad (7)$$

where the complex field is represented in the Madelung's form, $U \equiv |U|\exp(i\phi)$, and $k \equiv \partial\phi/\partial\xi$ is the respective wavenumber. The moments introduced in Eqs. (5), (6), and (7) determine the norm, $N$, momentum, $P$, and center-of-mass coordinate, $\bar{\xi} \equiv N^{-1}\int_{-\infty}^{+\infty} \xi |U|^2 d\xi$, of the wave packet.

## III. Analytical results

### A. Dynamics of moments

The system of exact evolution equations for the moments may be used for the derivation of approximate evolution equations for parameters of a soliton, see Refs. [49,50] and references therein. To this end, we adopt the usual ansatz for the moving soliton, with amplitude $A(t)$, wavenumber $k(t)$, and coordinate $\bar{\xi}$ defined above:

$$U(\xi,t) = A(t)\,\text{sech}\left[A(t)(\xi - \bar{\xi})\right]\exp\left[ik(t)\xi - \frac{i}{2}\int A^2(t)dt\right]. \tag{8}$$

The substitution of the ansatz into Eqs. (5)-(7) leads to the following evolution equations:

$$\frac{dk}{dt} = \frac{\beta}{3}A^2 - \frac{4}{15}\mu A^4, \quad \frac{dA}{dt} = \beta A k, \quad \frac{d\bar{\xi}}{dt} = -k, \tag{9}$$

which give rise to an obvious equilibrium state (alias fixed point, FP):

$$\mu_* \equiv 5\beta/(4A_0^2), \quad k_* = 0, \tag{10}$$

where $A_0$ is an arbitrary amplitude if the stationary soliton.

To analyze the evolution around the FP, we rescale the variables by defining $\tau \equiv t\beta A_0/\sqrt{6}$,

$$a \equiv A/A_0, \quad y \equiv k\sqrt{6}/A_0, \tag{11}$$

thus deriving a simple mechanical system from Eqs. (9):

$$\frac{dy}{d\tau} = 2a^2(1 - \lambda a^2), \quad \frac{da}{d\tau} = ay, \tag{12}$$

where $\lambda \equiv \mu/\mu_*$. Obviously, Eq. (12) conserves the corresponding Hamiltonian,

$$y^2 + \lambda(a^4 - 1) - 2(a^2 - 1) = y_0^2, \tag{13}$$

where $y_0$ is the value of $y$ at $a = 1$. Dynamical invariant (13) is drawn in the plane of $(y,a)$ in Fig. 1(a), for $y_0 = 0$ and different values of $\lambda$. Evidently, at $\lambda < 1$ (i.e., if the pseudo-SRS effect is relatively weak), the soliton's amplitude periodically oscillates between maximum and minimum values $a_{\max} \equiv A_{\max} / A_0 = \sqrt{(2-\lambda)/\lambda}$ and $a_{\min} = 1$ (the evolution of the amplitude is displayed in Fig. 2 for $\lambda = 1/4$). These values swap if the pseudo-SRS effect is stronger, *viz.*, $1 < \lambda < 2$ (the amplitude remains constant at $\lambda = 1$). As it follows from Eq. (12), oscillations of the soliton's amplitude translate into oscillations of its velocity, which are symmetric with respect to the positive and negative values.

Lastly, if the pseudo-SRS term is too large, with $\lambda \geq 2$, it destroys the soliton, as the evolution leads to the decay of the amplitude to $a = 0$, while the rescaled wavenumber takes the limit value $y_\infty \equiv \sqrt{\lambda - 2}$.

Further, at $y_0^2 > 0$ straightforward analysis of Eq. (13) demonstrates that the loop trajectories, which are seen in Fig. 1(a) for $y_0^2 = 0$, stretch in both positive and negative vertical directions (along the axis of $a$). In the same case, the critical value of the pseudo-SRS coefficient, which leads to the destruction of the soliton, decreases to $\lambda_{cr} = 2 - y_0^2$; thus, the solitons do not exist at all at $y_0^2 > 2$. Dynamical invariant (13) is schematically drawn in the plane of $(y, a)$ in Fig. 1(b), for $0 < y_0^2 < 2$ and different values of $\lambda$.

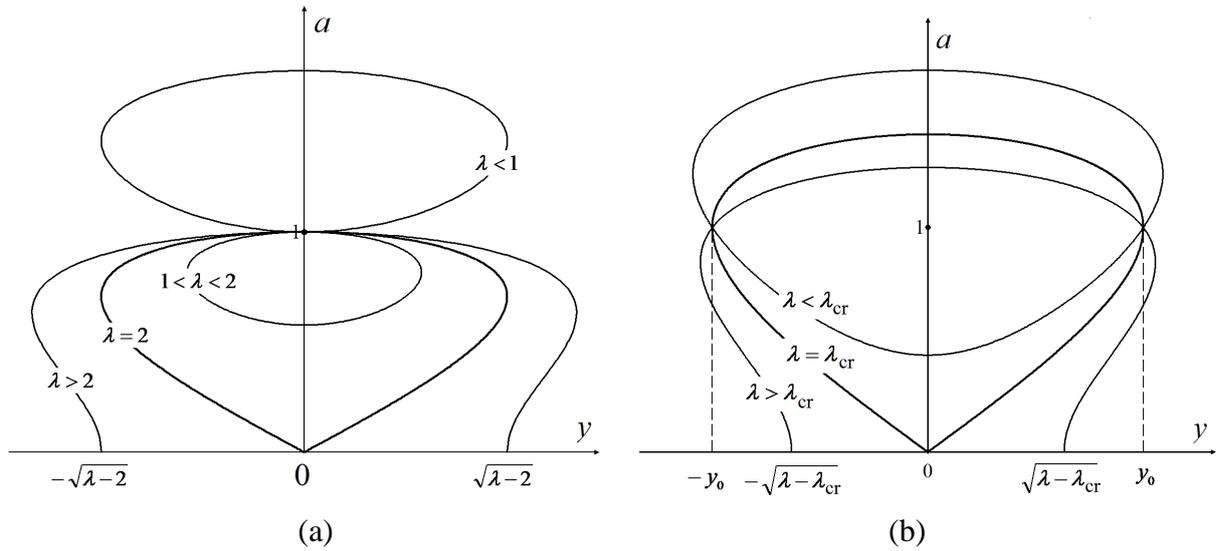

(a)          (b)

Fig. 1. Plots of dynamical invariant (12) in plane $(y, a)$ of the soliton's rescaled wavenumber and amplitude [see Eqs. (11)] for $y_0 = 0$ (a) and $0 < y_0^2 < 2$ (b), and different values of constant $\lambda$, see Eq. (28).

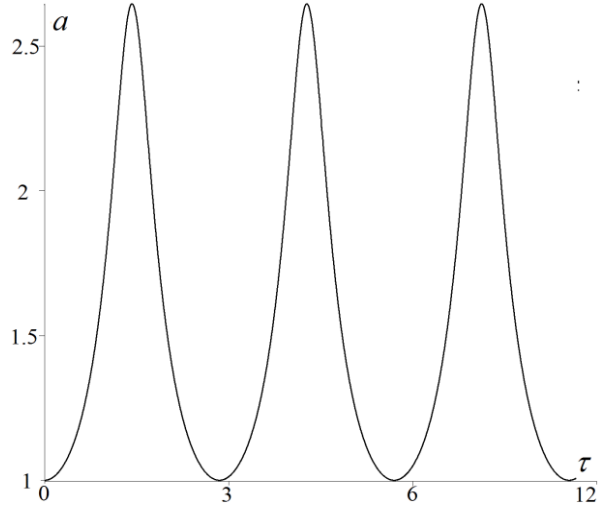

Fig.2. Time evolution of rescaled soliton's amplitude $a(\tau)$, as obtained from Eq. (12) for $\lambda = 1/4$ and $y_0 = 0$.

## B. The soliton solution

Stationary solutions to Eq. (3) are looked for as $U(\xi,t) = \psi(\xi)\exp(-i\Omega t)$, with a real soliton profile, $\psi(\xi)$, determined by the ordinary differential equation,

$$\frac{d^2\psi}{d\xi^2} + \beta\frac{d\psi}{d\xi} + 2\psi^3 - 2\Omega\psi - 2\mu\psi^2\frac{d\psi}{d\xi} = 0. \tag{14}$$

Next, assuming that the linear-gain and pseudo-SRS coefficients are small, $\beta, \mu \ll 1$, a solution to Eq. (14) is constructed in the perturbative form, $\psi = \psi_0 + \psi_1$, where $\psi_1$ is a small correction to $\psi_0$ determined by the linearization of the equation:

$$\frac{d^2\psi_0}{d\xi^2} + 2\psi_0^3 - 2\Omega\psi_0 = 0, \tag{15}$$

$$\frac{d^2\psi_1}{d\xi^2} + (6\psi_0^2 - 2\Omega)\psi_1 = -\beta\frac{d\psi_0}{d\xi} + 2\mu\psi_0^2\frac{d\psi_0}{d\xi}. \tag{16}$$

Equation (15) gives rise to the classical soliton, $\psi_0 = A_0 \operatorname{sech}(A_0\xi)$, with $\Omega \equiv A_0^2/2$. Then, substitutions $\eta \equiv A_0\xi$ and $\Psi \equiv \psi_1/\beta$ cast Eq. (16) in the form of

$$\frac{d^2\Psi}{d\eta^2} + \left(\frac{6}{\cosh^2\eta} - 1\right)\Psi = \frac{\sinh\eta}{\cosh^2\eta} - \frac{5}{2}\frac{\mu}{\mu_*}\frac{\sinh\eta}{\cosh^4\eta}. \tag{17}$$

At the FP, $\mu = \mu_*$ [see Eq. (10)], an exact solution to Eq. (17) can be found,

$$\Psi(\eta) = -\frac{\sinh\eta}{2\cosh^2\eta}\ln(\cosh\eta). \tag{18}$$

In terms of the original notation, the respective soliton solution to Eq. (4) is written as

$$U(\xi,t) = A_0\left[1 - \frac{\beta}{2}\tanh(A_0\xi)\ln(\cosh(A_0\xi))\right]\text{sech}(A_0\xi)\exp\left(-\frac{i}{2}A_0^2 t\right). \tag{19}$$

Note that the correction $\sim \beta$ breaks the spatial symmetry of the soliton. It is compared with the numerical solution in the next section.

## IV. Numerical results

### A. A single soliton

We simulated the evolution of wave packets governed by Eq. (3) with input $U(\xi, t=0) = \text{sech}\xi$ [which corresponds to $A(t=0)=1$ in Eq. (8)], for $\alpha = 1$ and, chiefly, for a fixed gain coefficient, $\beta = 1/10$, while the strength of the pseudo-SRS term, $\mu$, varied. Note that, with $A_0 = 1$, Eq. (10) yields the FP value $\mu_* = 1/8$. In direct simulations, the initial pulse for $\mu = 1/8$ is transformed into a stationary localized mode with zero wavenumber, which is very close to the analytical prediction, see Fig. 3(a). However, the relative correction $\sim \beta$ to the simplest sech profile in Eq. (19) is quite small, $\sim 0.03$, for $\beta = 1/10$, therefore, to better test the accuracy of the analytical approximation, in Fig. 3(b) we display the comparison of the analytical and numerically found profiles of stable stationary solitons for $\beta = 2/5$ [Eq. (10) yields $\mu_* = 1/2$ in this case]. The latter figure confirms the validity of the approximation based on Eq. (19).

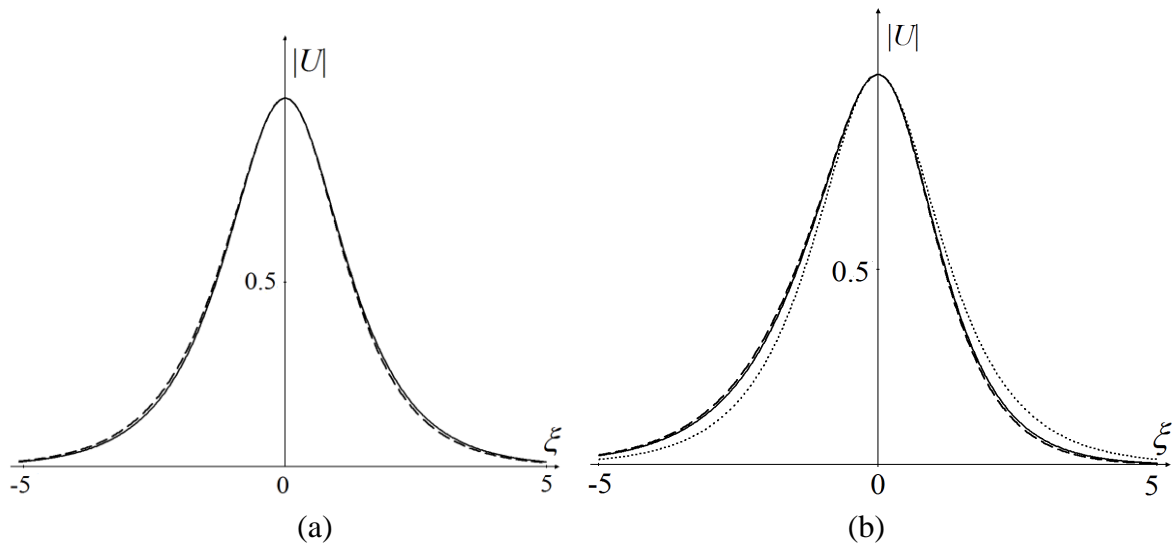

Fig. 3. (a) The solid curve is the result of the numerical solution of Eq. (3) for $|U(\xi)|$, which remains stationary in the time interval $5 < t \leq 200$, for $\beta = 1/10, \alpha = 1$ and the equilibrium value of the pseudo-SRS coefficient $\mu = 1/8 \equiv \mu_*$, as predicted by Eq. (10) for $A_0 = 1$. The dashed curve depicts $|U(\xi)|$, as predicted by analytical solution (19) for the same parameters. (b) The same for $\beta = 2/5, \alpha = 1$ and the respective equilibrium value $\mu_* = 1/2$, as predicted by Eq. (10) for $A_0 = 1$. For the sake of comparison, the dotted curve in panel (b) displays the unperturbed soliton profile, $U(\xi) = \text{sech}\,\xi$.

At values of the pseudo-SRS coefficient different from the FP value $\mu_*$, given by Eq. (10), the simulations produce nonstationary solitons, as shown in Fig. 4. In particular, in agreement with results of the above analysis (see Figs. 1 and 2), they remain robust oscillating pulses (stable against additional random perturbations), if $\mu$ is not too large [Figs. 4(a,b)], while larger values of $\mu$ lead to destruction of the soliton, as shown by Fig. 4(c). Boundary conditions with absorbers installed near edges of the integration domain were used in the simulations, to eliminate traveling perturbations driven by the convective instability.

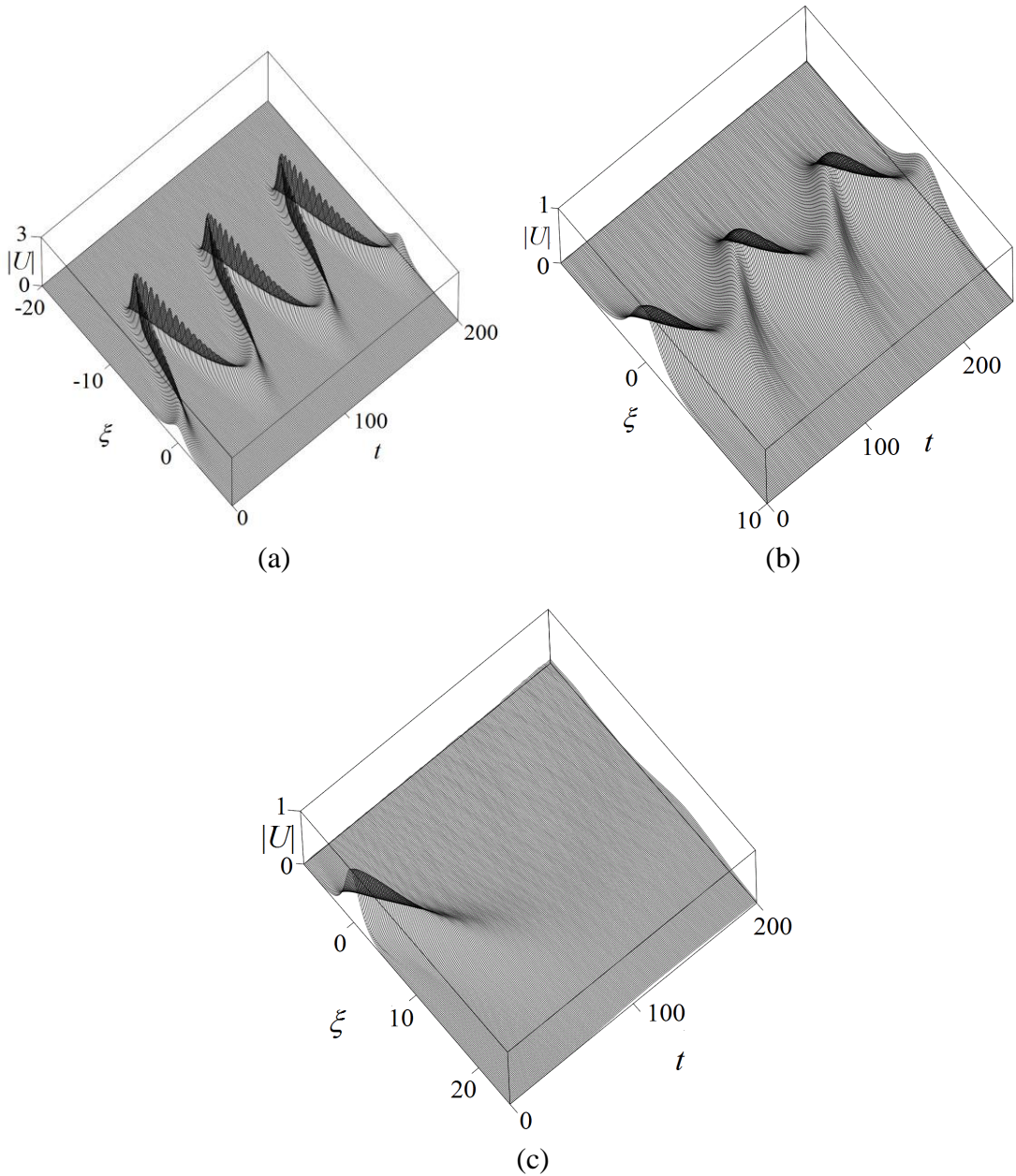

Fig.4. The numerically simulated evolution of the wave packets for $\beta =1/10$ and different strengths of the pseudo-SRS effect: (a) $\mu =1/32\equiv \mu_*/4$; (b) $\mu =1/6\equiv (3/2)\mu_*$; (c): $\mu =1/4\equiv 2\mu_*$, where $\mu_*$ is equilibrium value predicted by Eq. (10).

## B. Soliton-soliton interaction

The next natural step is to simulate interactions between solitons in the framework of Eq. (3). For this purpose, we solved the equation with input

$$U(\xi,t=0)=\text{sech}(\xi+\xi_0)\exp(i\varphi_0)+\text{sech}(\xi-\xi_0) \qquad (20)$$

at $\beta =1/10$, for different values of the pseudo-SRS coefficient, $\mu$, and initial separation between the solitons, $2\xi_0$. The phase shift is fixed to be $\varphi_0=\pi$, which corresponds, as usual, to the repulsive interaction between the solitons [1-7], while the attraction ($\phi_0=0$) makes the soliton pair unstable, leading to merger of the solitons. The evolution of $|U(\xi,t)|$, produced by the simulations for $\xi_0=4$, and different values of $\mu$, is shown in Fig. 5. In particular, at $\mu <\mu_*$, when a single soliton performs periodic oscillations [cf. Fig. 4(a)], the repulsive interaction between the solitons with the phase shift of $\pi$ causes the synchronization (phase locking) of their oscillations, as seen in Fig. 5(a). In the case of $\mu =\mu_*$, when single solitons maintain the stationary shape (see Fig.3), the interaction between the two of them gives rise to weak instability of the stationary pair, as shown in Fig. 5(b). Finally, if $\mu$ is taken too large, which implies destruction of an isolated soliton [see Fig. 4(c)], the interacting pair also suffers the destruction, although one soliton survives longer than the other, as shown in Fig. 5(c). Similar outcomes of the interaction were observed at other values of the parameters.

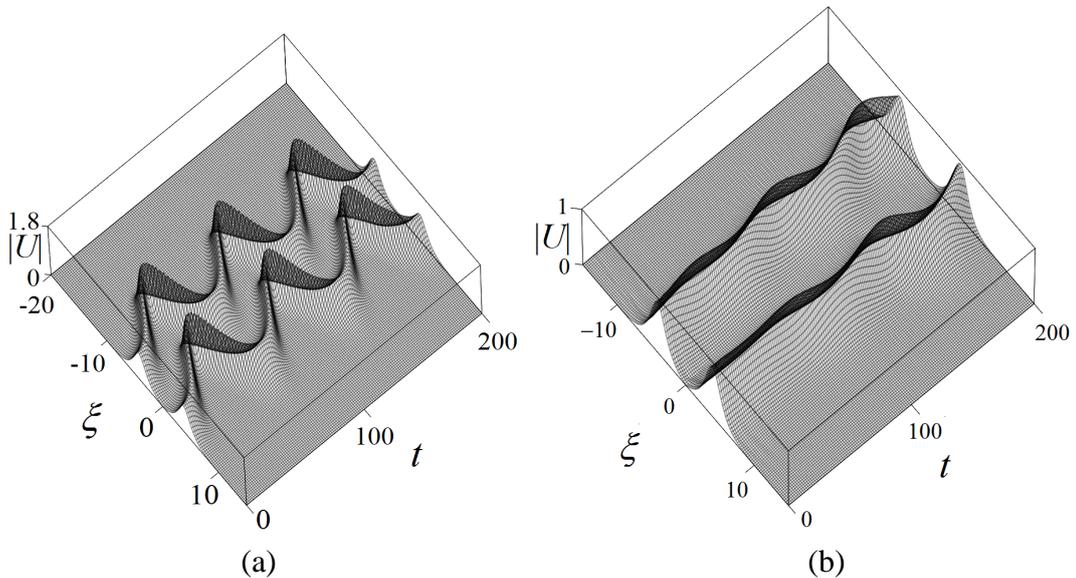

(a)        (b)

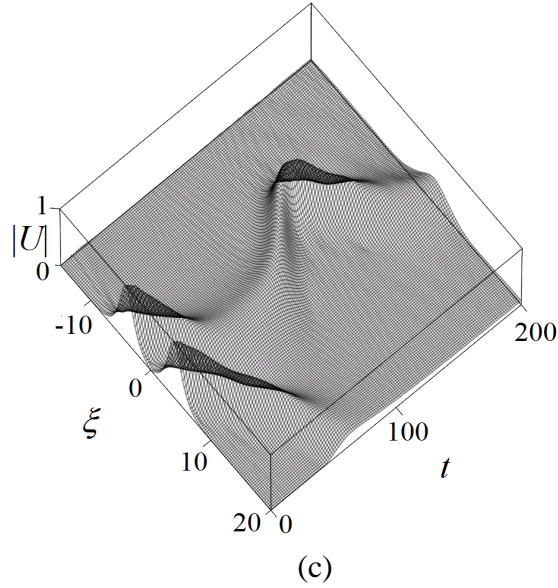

(c)

Fig.5. Numerically simulated evolution of soliton pairs, starting from input (20), for $\beta = 1/10$, $\xi_0 = 4$, $\varphi_0 = \pi$ and different strengths of the pseudo-SRS effect: (a) $\mu = 1/16$; (b) $\mu = 1/8$; (c): $\mu = 1/4$.

## 5. Conclusion

We have introduced the extended NLSE which includes the linear gain, represented by the first spatial derivative, and the pseudo-SRS (stimulated-Raman-scattering) term in the spatial domain. It appears as a natural model for wind-driven surface waves in the ocean, coupled to internal waves traveling at an interface between viscous fluid layers. Although the model gives rise to the inherent convective instability, it does not affect robust solitons, which are supported, in the static or dynamic (oscillatory) form, by the balance between the wavenumber upshift, induced by the linear gain, and the downshift, driven by the pseudo-SRS term. Both static and oscillating soliton states have been predicted by the analytical approximation and well corroborated by direct simulations. Solutions for the soliton pair with the phase shift of $\pi$ have been obtained in the numerical form, featuring the pair oscillating in the phase-locked form.

A natural extension of the present model may include higher-order terms, in the form of the cubic nonlinear and third-order linear dispersions. The dynamics of solitons in such a generalized model will be considered elsewhere.

## Acknowledgements

This work was done, in a part, in the framework of the Academic Fund Program at the National Research University - Higher School of Economics (HSE) in 2016-2017 (grant № 16-01-0002), and partly supported by a subsidy granted to the HSE by the Government of the Russian Federation for the implementation of the Global Competitiveness Program.